\documentclass[%
reprint,
superscriptaddress,
 amsmath,amssymb,
 aps,
pra,
]{revtex4-1}

\usepackage[english]{babel}
\usepackage[utf8x]{inputenc}
\usepackage[T1]{fontenc}
\usepackage{amsmath}
\usepackage{graphicx}
\usepackage[colorinlistoftodos]{todonotes}
\usepackage[colorlinks=true, allcolors=blue]{hyperref}
\usepackage{graphicx,amssymb,amsmath,amsthm,mathtools,mathrsfs,hyperref}
\usepackage{graphicx}
\usepackage{dcolumn}
\usepackage{bm}

\begin{document}

\preprint{APS/123-QED}

\title{
Quantum network security dependent on connection density between trusted nodes}

  
\author{Andrei Gaidash}
\affiliation{Leading Research Center ``National Center for Quantum Internet'',  ITMO University, 197101, 49 Kronverksky Pr., Saint Petersburg, Russia}
\affiliation{Department of Mathematical Methods for Quantum Technologies, Steklov Mathematical Institute of Russian Academy of Sciences, Moscow 119991, Russia}
\affiliation{Laboratory of Quantum Processes and Measurements, ITMO University,199034 Kadetskaya Line 3b, Saint Petersburg, Russia}
\affiliation{SMARTS-Quanttelecom LLC,6 line, Vasilievsky island, d.59, korp. 1, lit. B, Saint-Petersburg, 199178 Russia}

\author{George Miroshnichenko}
\affiliation{Waveguide Photonics Research Center, ITMO University, 197101, 49 Kronverksky Pr., Saint Petersburg, Russia}
\affiliation{Institute <<High School of Engineering>>, ITMO University, 197101, 49 Kronverksky Pr., Saint Petersburg, Russia}%

\author{Anton Kozubov}

  \email{avkozubov@itmo.ru}
  \affiliation{Leading Research Center ``National Center for Quantum Internet'',  ITMO University, 197101, 49 Kronverksky Pr., Saint Petersburg, Russia}
  \affiliation{Department of Mathematical Methods for Quantum Technologies, Steklov Mathematical Institute of Russian Academy of Sciences, Moscow 119991, Russia}
\affiliation{Laboratory of Quantum Processes and Measurements, ITMO University,199034 Kadetskaya Line 3b, Saint Petersburg, Russia}
\affiliation{SMARTS-Quanttelecom LLC,6 line, Vasilievsky island, d.59, korp. 1, lit. B, Saint-Petersburg, 199178 Russia}

\date{\today}

\begin{abstract}
In this paper we estimate how introduction of additional connections between trusted nodes through one, two and so on (i.e. connection density) to a quantum network with serial connection of trusted nodes affects its security. We provide proper scaling of failure probability of authentication and quantum key distribution protocols to the level of the whole quantum network. Expressions of the failure probability dependent on the total number of connected nodes between users and connection density for given mean failure probability of each element are derived. The result provides explicit trade-off between increase of the key transport security and consequent increase of spent resources. We believe that obtained result may be useful for both design of future network and optimization of existing ones.
\end{abstract}

\maketitle

\section{Introduction}

Quantum key distribution (QKD) is one of the most rapidly developed area of modern science. Crucial advantage of the technology is that security of private message transfer by quantumly distributed keys is based on the laws of quantum physics and not on the peculiar mathematical algorithms; the latter can be hacked in principle while one cannot trick fundamental laws of physics. Developing technology of QKD can be considered as a basis for future secure data transmission networks and global quantum internet as the final incarnation. The first steps towards the construction of quantum networks were presented in \cite{elliott2002building,elliott2005current,elliott2007darpa, poppe2008outline,dynes2019cambridge, peev2009secoqc,xu2009field,sasaki2011field,wang2014field}. Various network topologies were proposed and analyzed recently  as well as key transport schemes and its security estimation \cite{beals2008distributed,salvail2010security,barnett2011securing,phoenix2015relay,ma2017multiple, zhou2022quantum, rass2010building,solomons2021scalable,pattaranantakul2012secure}. However, the fundamental limitation on the distance between two neighbouring nodes forces the development of widespread quantum networks that cover big areas or elongated backbone networks that connects cities and countries. 

 Generally speaking there are two possible types of networks: with trusted and untrusted nodes. Untrusted nodes are usually based on some kind of quantum repeater \cite{jiang2009quantum,zhao2003experimental,wang2012quantum,ghalaii2020capacity} and it requires a quantum memory, for both it is hard to achieve necessary performance due to the current state of technology. However, there are special cases when only one untrusted node is required, one may utilize measurement-device-independent (MDI) \cite{lo2012measurement,braunstein2012side, tamaki2012phase,ma2012alternative,liu2013experimental,goodenough2021optimizing,ottaviani2019modular} QKD protocols. As an alternative to single-photon approach the first realization of twin-field (TF) QKD scheme with coherent states was proposed in \cite{lucamarini2018overcoming} which allows to overcome well-known fundamental limit of repeaterless quantum communications, i.e., the secret key capacity of the lossy communication channel \cite{pirandola2017fundamental} (also known as the Pirandola-Laurenza-Ottaviani-Banchi bound) \cite{pirandola2017fundamental,pirandola2018theory}. Moreover a bunch of new approaches for realization of TF QKD protocol were proposed in  \cite{minder2019experimental, wang2019beating, zhong2019proof,chistiakov2019feasibility}, as well as multiple users variant of twin-field like QKD \cite{grasselli2019conference}. 
 However even those implementations of multiple users variants of MDI or TF QKD systems (similar to the star network topology with untrusted node in the centre) are combined in a widespread network by trusted nodes, for instance see Fig.~1 in \cite{ottaviani2019modular}. So besides true quantum repeaters trusted node paradigm seems to be inevitable and one should consider configurations of trusted nodes and connections between them in order to estimate how probabilistic properties of each node are transferred to the level of the whole network, e.g. the most desired one is the security properties of networks and their key transport protocols.

The aim of this paper is to estimate how introduction of new connections to widely used serial connection (or increase of trusted nodes density in global quantum internet in the future) affects security of quantum network. We do not consider a limited amount of compromised nodes like in \cite{beals2008distributed,salvail2010security}. Our network segment configuration (meaning that we may consider chosen end-to-end path within wider network) and eavesdropping model are similar to one presented in \cite{barnett2011securing}. However, in our approach we do not monitor the presence of the eavesdropper in the nodes by dropping out any of the relays. One of the purposes of the approach is to estimate the mean probability of successful key transfer considering any possible configuration of compromised nodes and intercepted QKD links. Thus in this paper we demonstrate the appropriate key transfer technique and the general method for estimation of its successful implementation probability.

This paper is organized as follows. Section~\ref{keytrans} describes the topology of considered network segments and the key transfer protocol in details. In Section~\ref{attack} we provide the explicit description of the network security and its estimation. In Section~\ref{results} we discuss the obtained results.

\begin{figure}[tp]
\begin{center}
\includegraphics[width=1\linewidth]{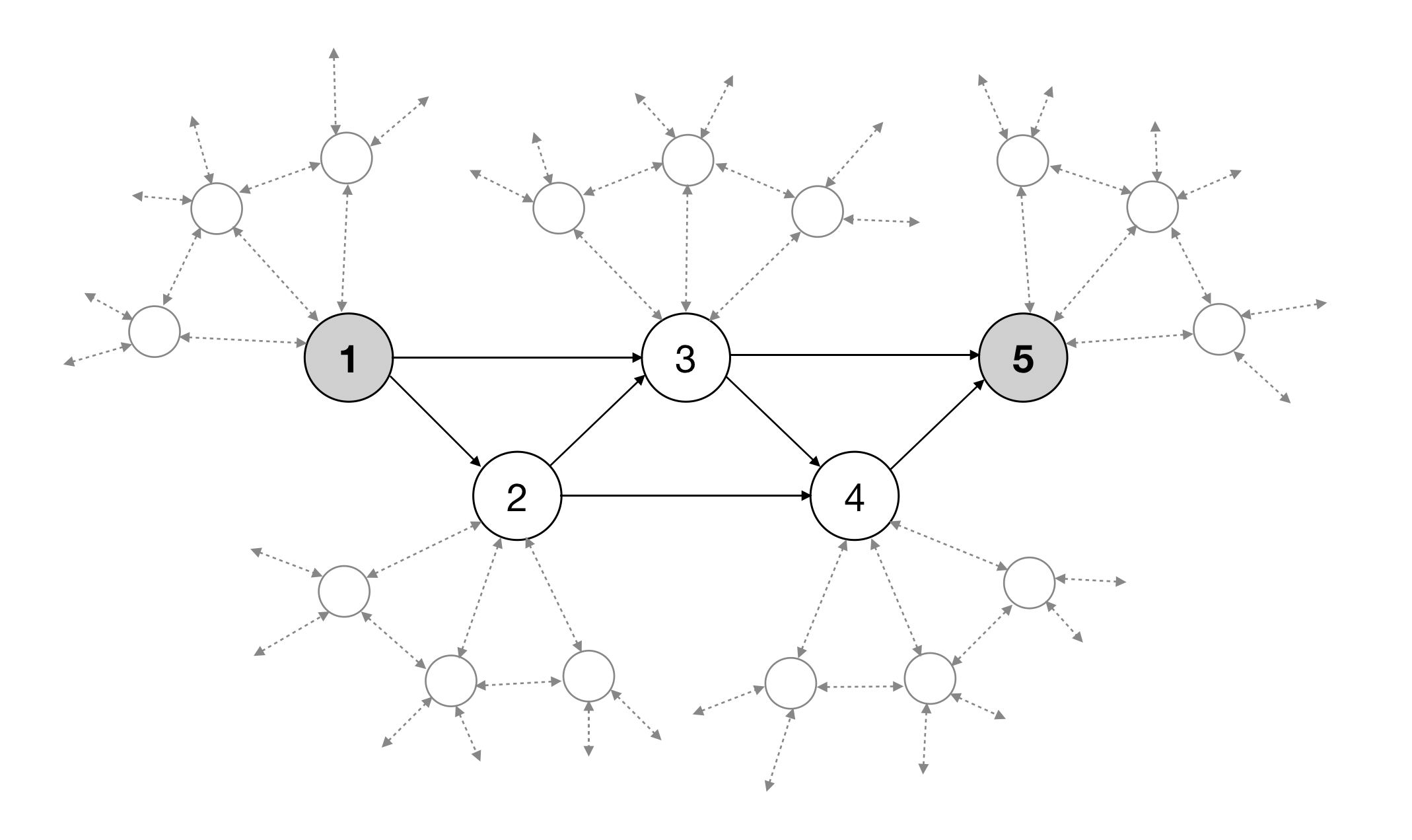} 
\end{center}
\caption{Visualization of particular segment of widespread quantum network. Circles are trusted nodes, connections between them are QKD links (also they are connected by classical channels that are not shown in the figure). Key transport is organized between grey-shaded nodes for a certain session.} 
\label{fig11}
\end{figure}

\section{Configuration and key transport protocol}\label{keytrans}
\begin{figure}[tp]
\begin{center}
\includegraphics[width=1\linewidth]{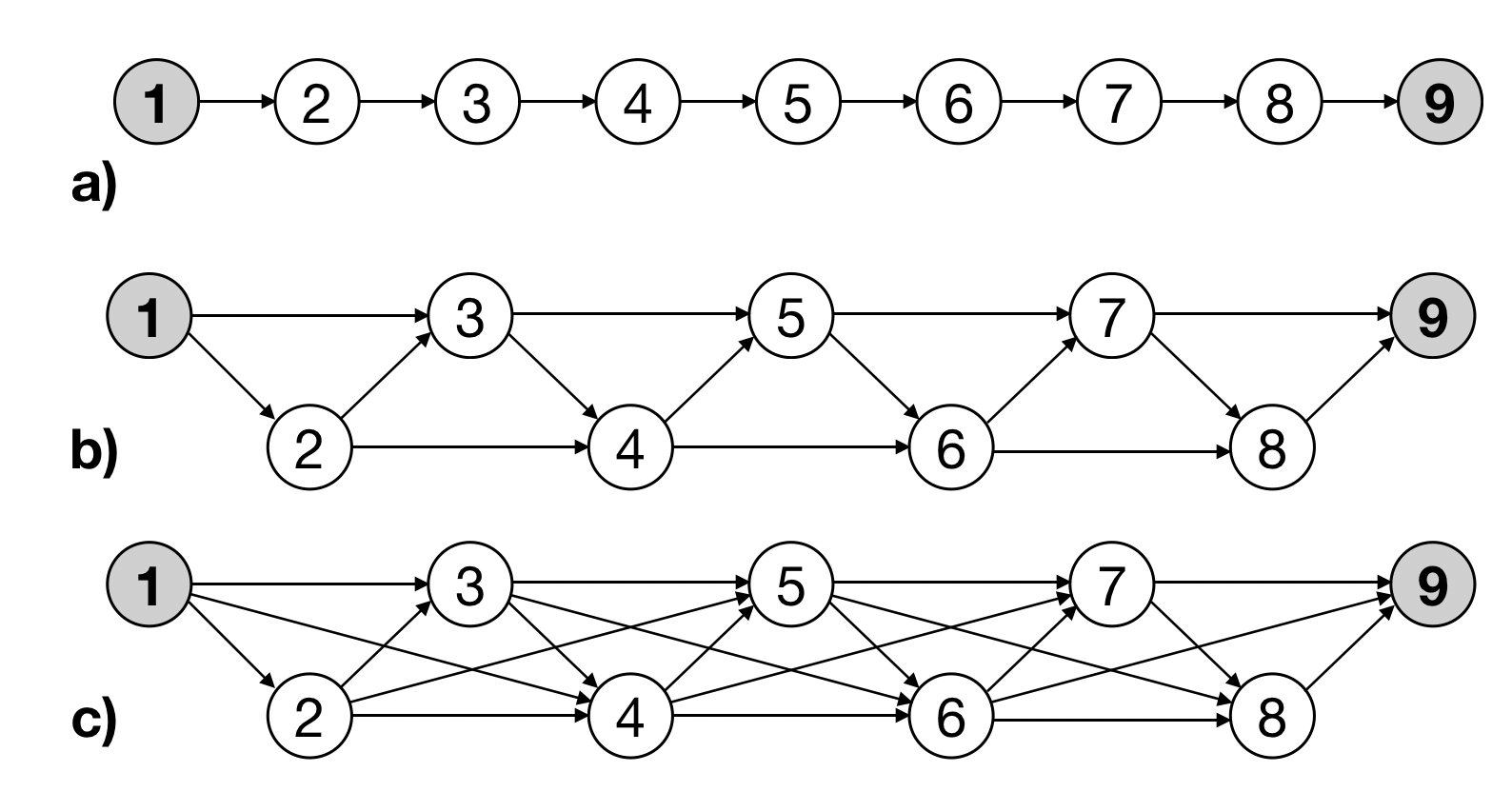} 
\end{center}
\caption{Visualization of quantum key distribution network segment (for a given key transport session between grey-shaded nodes) with different amount of additional connections, i.e. connection density. One may view the increase of connection as increased density of nodes and connections in the global quantum internet. Case with $N=9$ is considered as an example. a) Typical serial connection, $c=1$. b) Serial connection and additional connection though one node, $c=2$. c) Serial connection and additional connection though one and two nodes, $c=3$.} 
\label{fig1}
\end{figure}

In this paper we would like to consider a segment of a quantum network that connects two users within it (see Fig.~\ref{fig11}); considered part of a network contains $N$ nodes that at least serially connected to each other and may have additional connections through one, through two and up to through $c-1$ (see Fig.~\ref{fig1} for example), also total number of connections that require QKD links is $c(N-\frac{c+1}{2})$. Keys are distibuted quantumly between each pair of connected nodes. We assume that utilized quantum key distribution protocol is $\varepsilon_{qkd}$-secure, e.g. \cite{renner2008security}. Classical data encrypted by quantumly distributed keys is transferred in one direction (at least for a current session). The latter may be explicitly described by adjacency matrix $A$ that is matrix with $1$ at $k$-diagonals for $1\le k\le c$ and $0$ elsewhere. This configuration describes unidirectional connections between neighbouring nodes and through up to $c-1$ nodes.  Considered straightforward configuration of the network implies rather simple analysis and presence of useful properties. Also we believe that in principle properties of more complicated adjacency matrix configurations may be investigated by perturbation theory or other methods. However there is a high chance that particular segment in networks with dense node distribution can be described by adjacency matrix $A$ with symmetric properties as it noted earlier.
 
\begin{figure}[tp]
\begin{center}
\includegraphics[width=1\linewidth]{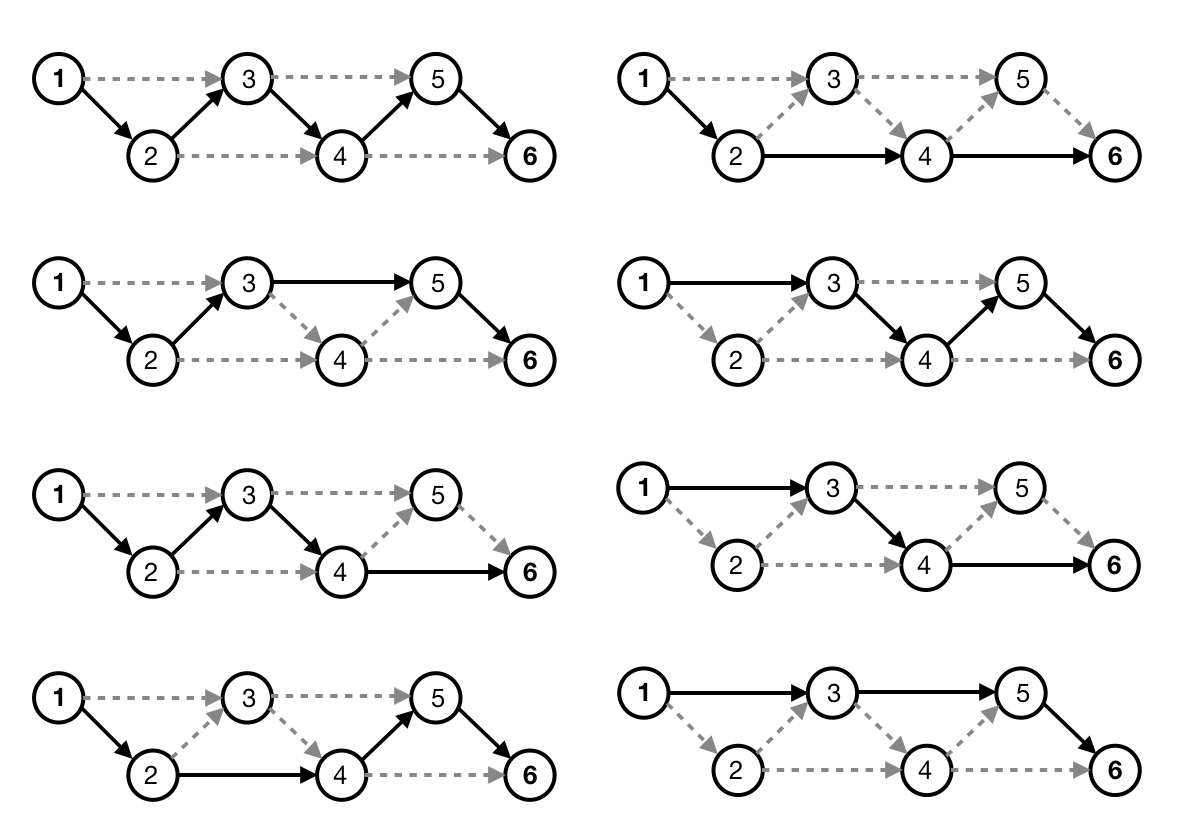}
\end{center}
\caption{Visualization of all possible key transport routes for quantum network segment with $N=6$ and $c=2$. Routes presented with the black solid lines, grey dashed lines present non-utilized (for a certain keys) connections. Total number of routes is $F^{(2)}_6=8$, i.e. sixth Fibonnacci number.}
\label{fig2}
\end{figure}

 In particular the total number of routes between users in a certain session is $F^{(c)}_N$, where the latter is $N^{th}$ $c-$annacci number (see Fig.~\ref{fig2} for example). Thus one may apply this property in order to construct key transport protocol similar to \cite{barnett2011securing,ma2017multiple}. Each route is assigned to transfer one of the keys $K_i$, where $1\le i\le F^{(c)}_N$. Quantumly distributed keys are used in order to transfer several $K_i$ with routing instructions as encrypted messages between the nodes. Then final key is $K=\bigoplus_{i}K_i$, where $\bigoplus$ is bitwise XOR operation. This method of key transport guarantees that compromising of one node does not reveal transferred key to an adversary. One may see App.~\ref{appex} where simple example of how key transport protocol works is considered. It should be noted that number of routes $F^{(c)}_N$ for large amount of nodes becomes enormous. This should be kept in mind and one may change routing scheme (e.g. decrease amount of routes to a certain degree). However it makes analysis intricate and we left this discussion beyond the scope of the paper.
 
Authentication protocols with failure probability $\varepsilon_{auth}$ are implemented in order to ensure that each node is trusted before QKD sessions; pool of preshared keys is used for this purposes, also it is updated with a part of quantumly distributed keys. It should be noted that authentication problem can be considered separately to QKD problem and then combined by composition principle \cite{portmann2014cryptographic}, thus we are eligible to assume some $\varepsilon_{qkd}$-security of the QKD protocol and do not consider it in details. Described key transport protocol succeeds if there is at least one route from the first node to the last one that goes only through authorized trusted nodes.

\section{Key transport security problem}\label{attack}

In the consideration of QKD network performance we utilize the next assumptions:
\begin{enumerate}
    \item Nodes are assumed to be trusted. Authentication protocol is assumed to work properly and fail with at most $\varepsilon_{auth}$ probability for each node. All nodes are attacked separately and simultaneously every key transport session. 
    \item QKD links are assumed to work properly between all nodes and be $\varepsilon_{qkd}$-secure. All QKD links are attacked separately and simultaneously every key transport session.
    \item Distance between two neighbouring nodes are less than the limiting one. Distance between the most distant directly connected (through $c-1$) nodes should be considered as the limiting one.
\end{enumerate}

The problems one faces in QKD network security estimation are pretty similar to the problems in point-to-point QKD links. Thus one has to deal with both attacks on authentication of nodes and the QKD protocol for every link. Basically failure probabilities of an authentication protocol ($\varepsilon_{auth}$) and a QKD protocol ($\varepsilon_{qkd}$) can be considered separately according to composition principle \cite{portmann2014cryptographic}. However, regarding network implementation we should correctly scale both notations considering certain restrictions.

\subsection{Authentication scaling problem}

\begin{figure}[tp]
\begin{center}
\includegraphics[width=1\linewidth]{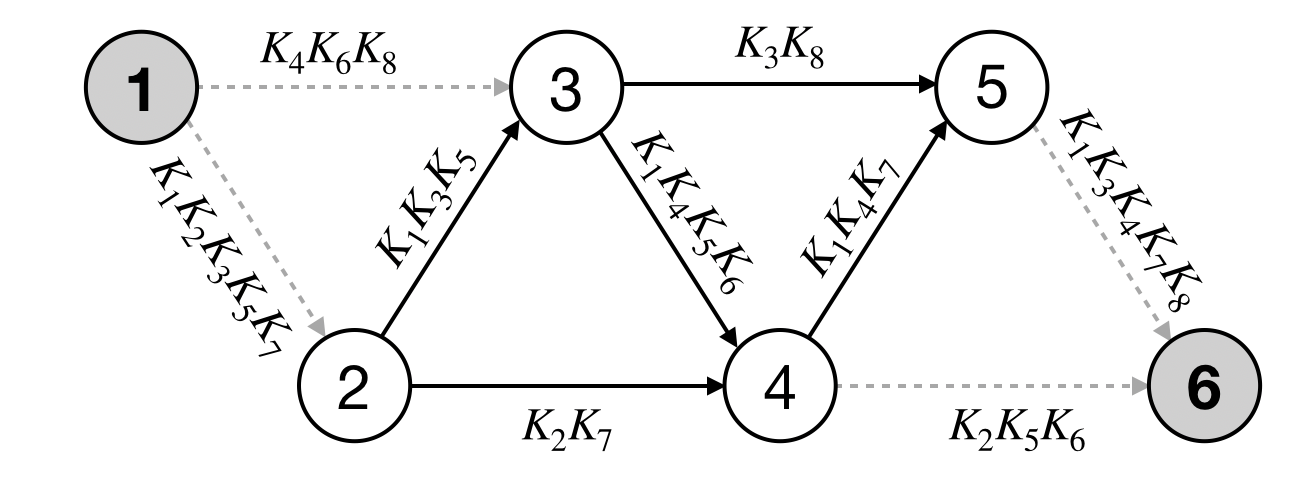}
\end{center}
\caption{Visualization of the routing scheme for key transport protocol of quantum network segment with $N=6$ and $c=2$. Each route is assigned to transfer one of the keys $K_i$, where $1\le i\le F^{(c)}_N$. Quantumly distributed keys are used in order to transfer several $K_i$ with routing instructions as encrypted messages between the nodes. Then final key is $K=\bigoplus_{i}K_i$, where $\bigoplus$ is bitwise XOR operation. Number of grey-shaded dashed links that goes out of the first node (their amount is $c$) or goes in the last node (also $c$) are the lowest number of links to be intercepted in order to obtain all $K_i$, other combinations have at least $c+1$ links.} 
\label{fig999}
\end{figure}

There is a possibility of ``man-in-the-middle'' attack, basically an attack on authentication protocol, when eavesdropper fully duplicates one of the nodes and acts like it and neighbouring nodes do not suspect anything. Then after QKD between each pair of nodes (including compromised nodes), at the moment when key transport with trusted nodes is performed, compromised node can obtain transferred through it information. Basically it is the attack on a classical key transport (with quantumly distributed keys) scheme and not on the QKD ``part'' of the network. In order to maximize the amount of compromised nodes eavesdropper should attack simultaneously all trusted nodes in a certain segment. The condition of the successful attack is to compromise at least $c$ nodes in a row (by the order of nodes in the longest path). This condition follows from the topology of network, since if $c$ nodes in a row are compromised there is no possible routes between the first and the last nodes that can be constructed. Then all $K_i$ can be known by eavesdropper. It also should be mentioned that eavesdropper cannot keep compromised node because knowledge of preshared key is required. Eavesdropper may attempt new attack at the beginning of each session.

One may consider full problem as it is solved in App.~\ref{app}. Alternatively, another way is consideration of approximated solution (i.e. the lowest order term, which works great with low probabilities) as follows: the lowest amount of compromised nodes for successful attack is $c$ and there is $N-c-1$ possible configurations of them to be located in a row within $N-2$ nodes (we do not consider compromising of the first and the last nodes). Then it is straightforward that overall probability of successful attack on authentication protocol is as follows:
\begin{gather}
    \varepsilon_{1}\approx (N-c-1)(\varepsilon_{auth})^c,\label{eps1}
\end{gather}
the latter approximation is reasonable for small $\varepsilon_{auth}\le \big(\frac{1}{N-c-1}\big)^{\frac{1}{c}}$ that can be easily satisfied.

\subsection{QKD security scaling problem}

In order to estimate network security one should also consider simultaneous attacks on all QKD links that may provide transferred key to an adversary. We utilize the lowest order term approximation as we did in previous subsection. Adversary needs to intercept quantum key at certain links in a way that all $K_i$ are transferred by those links, e.g. see Fig.~\ref{fig999}. It is obvious that the least amount of links that transfer all $K_i$ are those connected to the first or the last node, in either case there are $c$ links (higher terms are at least of order $c+1$). Hence successful attack on QKD links can be performed with the following probability:
\begin{gather}
  \varepsilon_{2}\approx 2(\varepsilon_{qkd})^c,\label{eps2}
\end{gather}
this approximation is reasonable for $c>1$ and $\varepsilon_{qkd}\le (\frac{1}{2})^\frac{1}{c}$ that can be easily satisfied. It should be noted that specifically in case $c=1$ the latter expression is $\varepsilon_{2}= (N-1)\cdot\varepsilon_{qkd}$.

\subsection{Quantum network security}

The main result of our paper is the security notation for arbitrary configurations of quantum networks utilizing key transport protocol. Following the composition principle we should bound failure probability of quantum network, $\varepsilon_{qn}$, via scaled failure probabilities of authentication, $\varepsilon_{1}$, and QKD protocols,  $\varepsilon_{2}$, using next equation: 
\begin{gather}
    \varepsilon_{qn}=\varepsilon_{1}+\varepsilon_{2}\approx(N-c-1)(\varepsilon_{auth})^c+ 2(\varepsilon_{qkd})^c.\label{epsqn}
\end{gather}

\section{results and discussion}\label{results}

The obtained results can be applied to two different scenarios. On one hand one may consider obtained results regarding current state of the art. Quantum networks are in the early stages of development so one can design its configuration dependent on different purposes. As it is shown in Eqs.~\ref{eps1} and~\ref{eps2} introduction of additional connections reduces the probabilities of the attack as power function ($\varepsilon_{auth}\rightarrow (\varepsilon_{auth})^c$ and $\varepsilon_{qkd}\rightarrow (\varepsilon_{qkd})^c$). However at the same time requirements for maximal allowed for QKD losses in the quantum channel are increased ($\eta\rightarrow\eta^c$) or a number of nodes per maximal allowed distance is increased in $c$ times (the cost of $N$ nodes $\rightarrow$ the cost of $N\cdot c$ nodes), also total number of edges (i.e. number of QKD connections) are increased as well ($N-1\rightarrow c(N-\frac{c+1}{2})$). At the same time one should avoid enormous number of routes in the key transport scheme or be capable to provide fast enough data transfer rates. Security of the quantum network is in priority however one should make the decision about the latter trade-off. The result provides simple dependencies on $c$ parameter (where it can be considered as density of connections in the network) in order to make analysis of the trade-off as easy as possible. Thus one can set the topology (and optimize the cost) of the network at the stage of its design in order to achieve necessary security.

On the other hand we may consider obtained results regarding future global quantum internet where there are already dense trusted nodes distribution within it. In this case one may utilize obtained results in order to adjust parameters of particular key transport session considering necessary security provided by minimal spent resources. More specifically, if universal hash functions \cite{carter1979universal} are used in order to authenticate users then probability of node to be compromised is of the order of hash function collision, i.e. $2^{-n/2}$, where $n$ is the length of hash output. Then according to Eq.~\ref{eps1} hash output length may be reduced by a factor of $c\log_{N-2}(N-c-1)$ while we preserve overall security. Optimal value of $c$ that reduces hash output at most can be obtained numerically by solving the following equation:
\begin{gather}
    (N-c-1)\ln(N-c-1)=c,
\end{gather}
where one should keep in mind that $1\le c < N-2$ and it is integer. Approximate solution for the latter equation can be found as $c\approx \frac{(N-1)\ln (N-1)}{\ln (N-1)+2}$. We believe this may be useful in the context of key recycling paradigm \cite{portmann2014key}.

Another point of view on the problem is that one may estimate network failure probability considering $\varepsilon_{auth}$ as mean nodes' failure probability. Then obtained result in Eq.~\ref{eps1} shows the probability that there will be no working routes that connect the first and the last nodes, i.e. overall denial of service probability.

At the end of the day we want to conclude that obtained expression in Eq.~\ref{epsqn} may be useful for both design of future networks and optimization of existing ones.

\section*{acknowledgements}
The work was done by Leading Research Center ``National Center of Quantum Internet'' of ITMO University during the implementation of the government support program, with the financial support of Ministry of Digital Development, Communications and Mass Media of the Russian Federation and RVC JSC; Grant Agreement ID: 0000000007119P190002, agreement No. 006-20 dated 27.03.2020.

\appendix
\section*{Appendix}

\section{Key transport protocol example}\label{appex}

Let us consider simple example of how key transport protocol works for $N=6$ and one additional connection, i.e. $c=2$. The task is to securely transfer massage $M$ from the first node to the last one (and there is no direct QKD connection between them) following described below algorithm:

\begin{enumerate}
    \item Each node authenticates in the network;
    \item Between each pair of nodes QKD is performed, key $k_{12}$ is shared between the first and the second node, key $k_{13}$ is shared between the first and the third node and so on, i.e. key $k_{ij}$ is shared between $i^{th}$ and $j^{th}$ nodes if there is QKD link between them;
    \item Network defines (it can be done by software defined network (SDN) principles) total number of routes for key transport $F^{(c)}_N$, in our case it is $F^{(2)}_6=8$, they can be observed in Fig.~\ref{fig2};
    \item For each route the first node generates $K_i$ with $1\le i \le 8$. Then the first node (or network itself by SDN principles) develop routing scheme $R$ as it is shown in Fig.~\ref{fig999} and send it to other nodes;
    \item The first node transfers encrypted messages $(K_1 K_2 K_3 K_5 K_7) \bigoplus k_{12}$ and $(K_4 K_6 K_8) \bigoplus k_{13}$ to the second and the third nodes correspondingly by open classical channel (OCC, e.g. the Internet), where by $K_iK_j$ we assume concatenated bitstrings $K_i$ and $K_j$, $\bigoplus$ is bitwise XOR;
    \item The second node decrypts obtained message by applying known quantum key as $K_1 K_2 K_3 K_5 K_7=(K_1 K_2 K_3 K_5 K_7) \bigoplus k_{12}\bigoplus k_{12}$ and splits it in according to the routing scheme $R$ by $K_1 K_3 K_5$ and $K_2 K_7$;
    \item The second node sends $(K_1 K_3 K_5) \bigoplus k_{23}$ and $(K_2 K_7)\bigoplus k_{24}$ to the third and fourth nodes correspondingly by OCC;
    \item And so on, following eight messages are sent by OCC during the session:
    \begin{itemize}
        \item $(K_1 K_2 K_3 K_5 K_7) \bigoplus k_{12}$,
        \item $(K_4 K_6 K_8) \bigoplus k_{13}$,
        \item $(K_1 K_3 K_5 ) \bigoplus k_{23}$,
        \item $( K_2 K_7) \bigoplus k_{24}$,
        \item $(K_1 K_4 K_5 K_6) \bigoplus k_{34}$,
        \item $(K_3 K_8) \bigoplus k_{12}$,
        \item $(K_1 K_4 K_7) \bigoplus k_{45}$,
        \item $(K_2 K_5 K_6 ) \bigoplus k_{46}$,
        \item $(K_1 K_3 K_4 K_7 K_8) \bigoplus k_{56}$;
    \end{itemize}
    \item By doing so the last node obtains all $K_1,...,K_8$. Then the first and the last nodes obtain encryption key $K=\bigoplus_{i}K_i$ known only to them;
    \item The first node encrypts message $M$ as $M\bigoplus K$ and transfers it by OCC to the last node where one decrypts the message by $M=M\bigoplus K\bigoplus K$.
\end{enumerate}

\section{Strict derivation}\label{app}
To estimate the probability of the successful attack one should calculate the ratio between the number of all possible combinations of compromised nodes leading to the complete key eavesdropping and the number of all possible combinations of compromised nodes.

Let us consider the estimation algorithm in more details:
\begin{enumerate}
    \item Eavesdropper attacks every trusted node (in particular segment) with the mean success probability $p$.
    \item According to the Bernoulli scheme the probability of compromising $m$ nodes is as follows:
    \begin{gather}
        p_m=\binom{N-2}{m}p^m(1-p)^{N-m-2},
    \end{gather}
    where $\binom{a }{b}=\frac{a!}{b!(a-b)!}$ is the corresponding binomial coefficient. 
    We consider $N-2$ nodes since the first and the last node are assumed to be not under the attack.
    \item Since the probability of compromising $m$ nodes is known, one should finally estimate the amount of compromised nodes combinations leading to a successful attack. Necessary combinations are where at least $c$ in a row nodes (by their order) are compromised; it is as follows:
    \begin{gather}
        f(N,m,c)=\binom{N-2}{m}-\text{cf}\Big[\big(\sum_{k=0}^{c-1}x^k\big)^{N-m-1}\Big]_{m},\label{fnmc}
    \end{gather}
    where $\text{cf}[\ \cdot\ ]_{m}$ is corresponding coefficient of $x^{m}$ summand. The latter expression is obtained heuristically by observation of the result of numerical simulations. However, alternatively the expression can be derived in different way as follows:
    \begin{gather}
         f(N,m,c)=\sum_{j=1}^{\left\lfloor{\frac{m}{c}}\right\rfloor}(-1)^{j+1}\binom{N-m-1}{j}\binom{N-2-c\cdot j}{m-c\cdot j},\label{fnmc1}
    \end{gather}

    \begin{figure}[tp]
\begin{center}
\includegraphics[width=1\linewidth]{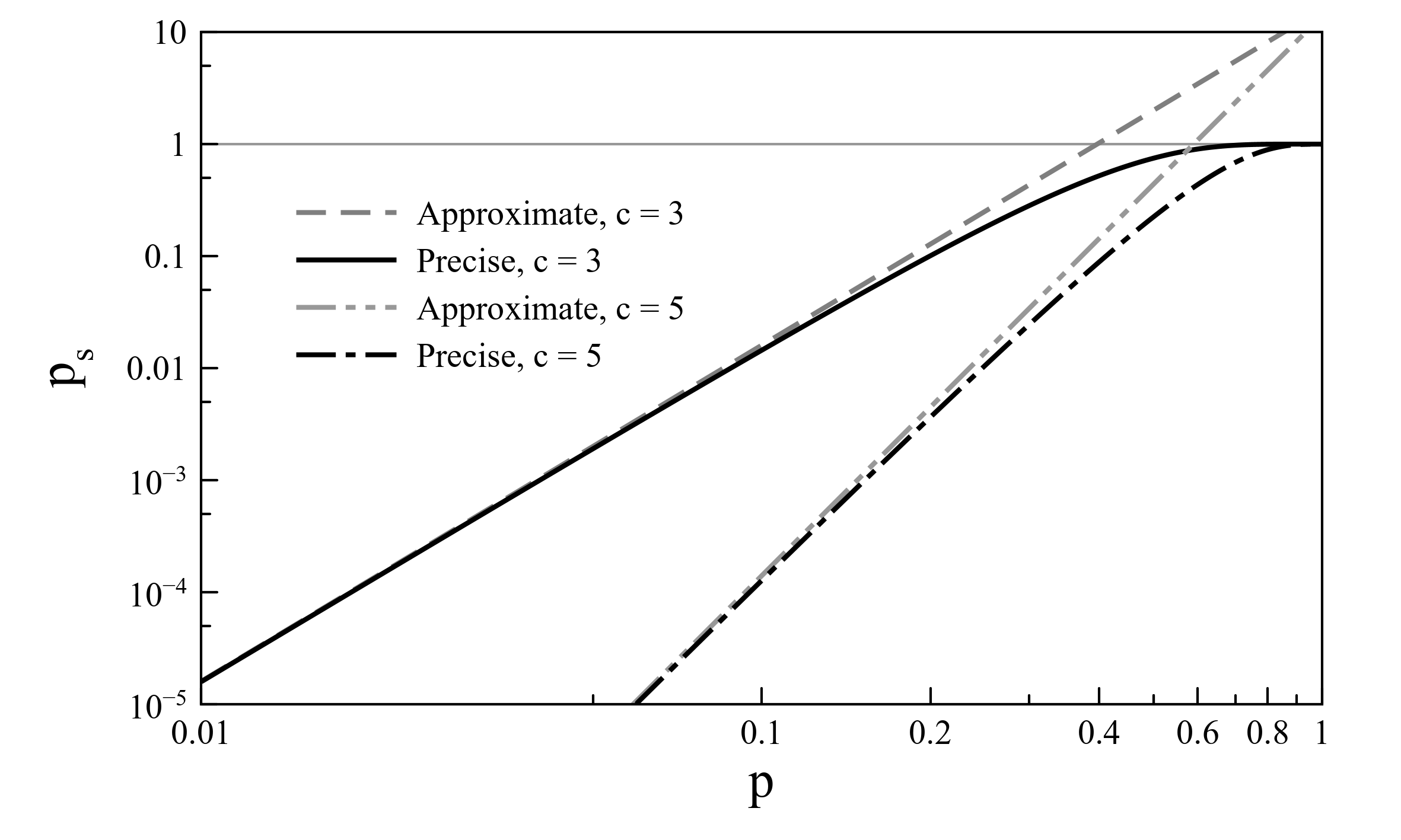}
\end{center}
\caption{Representative dependencies of successful attack on the network segment probability $p_s$ on mean probability $p$ to take over a node by the eavesdropper. Approximate (Eq.~\ref{psapprox}) and precise (Eq.~\ref{attackprob}) dependencies are shown. Two cases are considered: $c=3$ and $c=5$; $N=20$ as an example.} 
\label{fig789}
\end{figure}
    
    where $\left\lfloor{\ \cdot \ }\right\rfloor$ is floor function. More details on derivation of Eq.~\ref{fnmc1} as well as its equivalence to Eqs.~\ref{fnmc} are shown in the following Appendices correspondingly. Visualisation example of $f(N,m,c)$ is shown in Fig.~\ref{fig3}. Relation
    \begin{gather}
        p(s|m)=\frac{f(N,m,c)}{\binom{N-2}{m}}
    \end{gather}
    is conditional probability of successful attack when $m$ nodes are compromised.
    \item Then probability of successful attack is defined as follows:
    \begin{gather}
        p_{s}=\sum_{m=0}^{N-2}p(s|m)\cdot p_m,\label{attackprob}\\
        p_{s}\approx (N-c-1) p^c,\label{psapprox}
    \end{gather}
    the latter approximation is reasonable for small mean success probabilities $p\le \big(\frac{1}{N-c-1}\big)^{\frac{1}{c}}$. For $ p\ge \big(\frac{1}{N-c-1}\big)^{\frac{1}{c}}$ probability of successful attack $p_s$ is close to one. Behaviour of found precise and approximate expressions can be observed in Fig.~\ref{fig789}. As one may observe approximated result is the same as in the main body of the article.
\end{enumerate}

    \begin{figure*}[tp]
\begin{center}
{\includegraphics[width=0.9\linewidth]{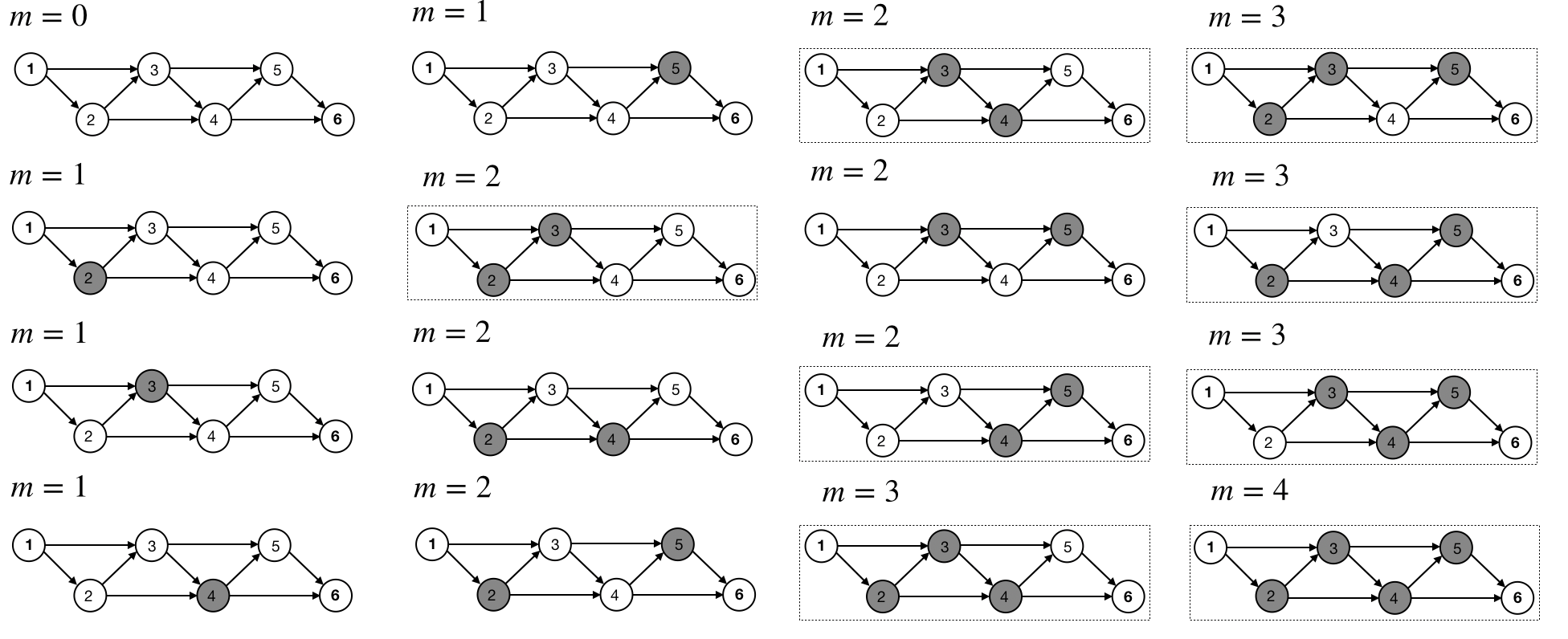}} 
\end{center}
\caption{Various configurations of the networ segment with $m$ compromised (colored with grey) nodes. Case with $N=6$ is considered as an example. Configurations where there is no path from the first node to the last one through uncompromised nodes are circled with dashed line. One may notice that circled configurations contain at least two nodes in a row (by the order of the nodes). The number of circled configurations for fixed $m$ is in accordance with Eq.~\ref{fnmc}.} 
\label{fig3}
\end{figure*} 

\section{Derivation of f(N,m,c) expression}\label{appb}
The problem is to define the number of combinations where $m$ entities (compromised nodes) are randomly distributed between $N-2$ positions (total number of trusted nodes in considered segment) and at least $c$ of them are located in neighbouring positions (at least $c$ in a row). Obviously, when $m<c$ there are no described combinations. Reasonable approach is to fix some $k\ge c$ neighbouring positions and observe the number of combinations of the rest $m-k$ located in the rest of $N-2-k$ positions. However in that case one should avoid counting multiple times the same configurations for different values of considered $k$. Thus it is necessary to follow the algorithm described below:
\begin{enumerate}
    \item Consider $c\le m <2c$. Step $i=0$. Total number of outcomes when $m$ occupied positions are neighbouring is $V(i=0)=N-m-1$, it is shown in Fig.~\ref{fig4} a).
    \item Step $i=1$. Consider $m-1$ neighbouring positions. Also we would like to prohibit occupation of the closest two positions (in order to avoid multiple counting of the same pattern), they are highlighted in Fig.~\ref{fig4} b) with light grey. Then there are $(m-1)+2$ ``occupied''(i.e. occupied and prohibited) positions that may ``touch'' the left and the right edges and $N-2-(m-1)-2$ vacant positions. Total number of outcomes is
    \begin{gather}
        V(i=1)=(N-2-(m-1)-2+1)\times\\
        \times\binom{N-2-(m-1)-2}{1}.\nonumber
    \end{gather}
    
    \begin{figure}[t]
\begin{center}
\includegraphics[width=0.9\linewidth]{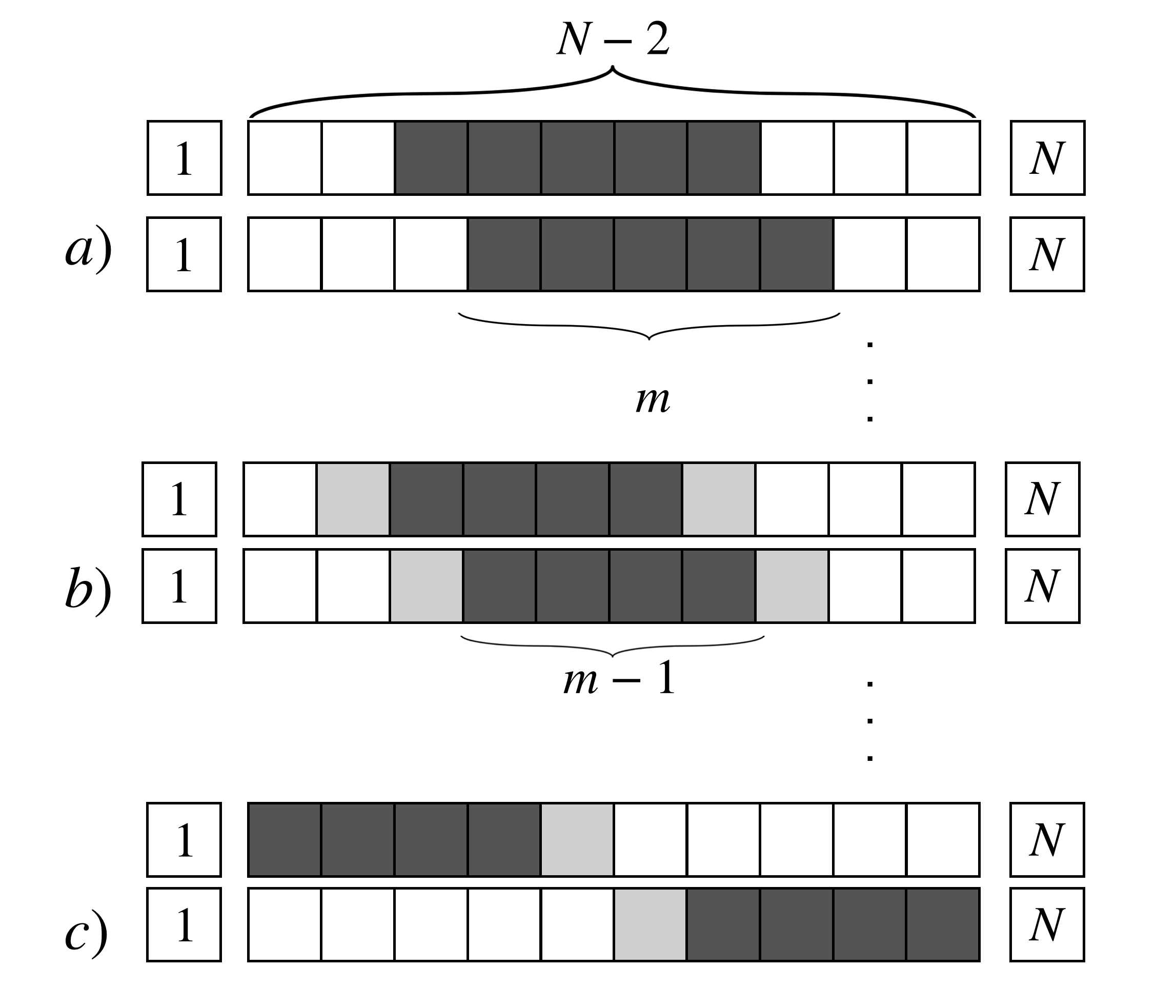}
\end{center}
\caption{Visualisation of the algorithm. Occupied positions is denoted by dark grey. Prohibited for occupation positions is denoted by light grey. a) Step $i=0$. Total number of outcomes when $m$ occupied positions are neighbouring is $V(0)=N-m-1$. b) Step $i=1$. Consider $m-1$ neighbouring positions, and occupation of the closest two positions is prohibited. Then there are $(m-1)+2$ ``occupied'' (i.e. occupied and prohibited) positions that may ``touch'' the left and the right edges and $N-2-(m-1)-2$ vacant positions. Total number of outcomes is $V(1)=(N-2-(m-1)-2+1)\binom{N-2-(m-1)-2}{1}$. c) Also $m-1$ occupied positions but they touch one of the edges. Occupation of the closest position is prohibited as well. Then the number of ``occupied'' positions is $(m-1)+1$ and the number of vacant positions is $N-2-(m-1)-1$; the number of combinations for considered $m$ is $W(1)=\binom{N-2-(m-1)-1}{1}$} 
\label{fig4}
\end{figure}

    \item Step $i\le m-c$. Consider $m-i$ neighbouring positions. We prohibit occupation of the closest two positions as well. Then there are $(m-i)+2$ ``occupied'' (i.e. occupied and prohibited) positions that may ``touch'' the left and the right edges and $N-2-(m-i)-2$ vacant positions. Total number of outcomes is
    \begin{gather}
        V(i)=(N-2-(m-i)-2+1)\times\\
        \times\binom{N-2-(m-i)-2}{i}.\nonumber
    \end{gather}
    \item Total number of combinations is as follows:
    \begin{gather}
        \sum_{i=1}^{m-c}V(i)= \sum_{i=1}^{m-c}(N-m-3+i)\binom{N-4-m+i}{i}.
    \end{gather}
    It should be noted that summation here starts with $i=1$ since when $i=0$ occupied positions touch edges. This case is considered further separately.
    \item Consider edges as it is shown if Fig.~\ref{fig4} c). Step $i\le m-c$. Then $m-i$ occupied positions touch one of the edges. We prohibit occupation of the closest position as well. Then the number of ``occupied'' positions is $(m-i)+1$ and the number of vacant positions is $N-2-(m-i)-1$; the number of combinations for considered $m$ is as follows:
    \begin{gather}
        W(i)=\binom{N-2-(m-i)-1}{i}.
    \end{gather}
    \item Total number of ``edge'' combinations is as follows:
    \begin{gather}
         2\sum_{i=1}^{m-c}W(i)= 2\sum_{i=1}^{m-c}\binom{N-3-m+i}{i},
    \end{gather}
    where factor of two is due to two edges.
    \item Finally the total number of allocation combinations is as follows:
    \begin{gather}
     V(0)+\sum_{i=1}^{m-c}\big(V(i)+2W(i)\big)=(N-m-1)\times\\
     \times\binom{N-2-c}{m-c}\nonumber
    \end{gather}
    where we utilize the following property:
    \begin{gather}
     \sum_{k=0}^{a}\binom{b+k}{k}=\binom{b+a+1}{a}.
    \end{gather}
    \item Then one should consider similar to previous steps for $2c\le m <3c$, $3c\le m <4c$ and so on. However at these steps one should be aware of possible double counting of some combinations (this can explain obtained further change of signs at summation). Presence of heuristic Eq.~\ref{fnmc} as the reference helps us to consider intricate avoiding of double counting at these steps in a right way. At the end of the day we obtain final expression for estimation of the number of combinations where $m$ entities (compromised nodes) are randomly distributed between $N-2$ positions (total number of trusted nodes) and at least $c$ of them are located in neighbouring positions (at least $c$ in a row) as follows:
    \begin{gather}
         f(N,m,c)=\sum_{j=1}^{\left\lfloor{\frac{m}{c}}\right\rfloor}(-1)^{j+1}\binom{N-m-1}{j}\times\\
         \times\binom{N-2-c\cdot j}{m-c\cdot j}.\nonumber
    \end{gather}
\end{enumerate}

\section{Equivalence of approaches}\label{appb}

In this section we consider the equivalence of heuristically obtained expression involving generating function in  Eq.~\ref{fnmc} and alternative expression in Eq.~\ref{fnmc1}. To do so lets consider the following steps:

\begin{enumerate}
    \item Let us consider the generating function from Eq.~\ref{fnmc}:

    \begin{gather}
    \sum_{i=0}^{c-1}x^i=\frac{1-x^c}{1-x}.
    \end{gather}
\item The next step is to raise it to the power $K=n-m-1$ and expand as folllows:
    \begin{gather}
    (1-x^c)^K\cdot(1-x)^{-K},\\
    (1-x^c)^K=\sum_{s=0}^K(-1)^s\binom{K}{s}\cdot x^{c\cdot s},\\
    (1-x)^{-K}=\sum_{q=0}^\infty x^N\cdot\binom{K+q-1}{q}.
    \end{gather}
\item Let us differentiate the expression:
    \begin{gather}
    (U(x)\cdot V(x))^{[m]}=\sum_{r=0}^{m}\binom{m}{r} U(x)^{[r]}  V(x)^{[m-r]},\\
      \Big((1-x)^{-K}\Big)^{[m-r]}\bigg|_{x=0}= \binom{K+m-r-1}{m-r}\cdot (m-r)!,\\
   \Big((1-x^c)^{K}\Big)^{[r]}\bigg|_{x=0}=(-1)^{\frac{r}{c}}\cdot\binom{K}{\frac{r}{c}}\cdot r!,\label{10}
    \end{gather}
 where $(\ \cdot \ )^{[m]}=\frac{d^m}{d x^m}$.
 Eq.~\ref{10} is as follows if $\frac{r}{c}$ is integer, otherwise it is equal to zero.
\item Lets substitute $r=j\cdot c$ and $K=N-m-1$:
    \begin{gather}
      \left[\left(\sum_{i=0}^{c-1}x^i\right)^{N-m-1}\right]^{[m]}\Bigg|_{x=0}=\\
       =m!\cdot\sum_{j=0}^{\left\lfloor{\ \frac{m}{c} \ }\right\rfloor} (-1)^j\cdot\binom{N-m-1}{j}\cdot\binom{N-2-c\cdot j}{m-c\cdot j}\nonumber\label{11}
    \end{gather}
\item And the final step is as follows:
    \begin{gather}
        \binom{N-2}{m}-\text{cf} \left[\left(\sum_{i=0}^{c-1}x^i\right)^{N-m-1}\right]_m=\\
        =\sum_{j=1}^{\left\lfloor{\ \frac{m}{c} \ }\right\rfloor} (-1)^{j+1}\cdot\binom{N-m-1}{j}\cdot\binom{N-2-c\cdot j}{m-c\cdot j}.\nonumber
    \end{gather}
\end{enumerate}

\bibliography{sample}

\end{document}